\documentclass[conference]{IEEEtran}
\IEEEoverridecommandlockouts
\usepackage{cite}
\usepackage{amsmath,amssymb,amsfonts}
\usepackage{algorithmic}
\usepackage{graphicx}
\usepackage{textcomp}
\usepackage{xcolor}
\def\BibTeX{{\rm B\kern-.05em{\sc i\kern-.025em b}\kern-.08em
    T\kern-.1667em\lower.7ex\hbox{E}\kern-.125emX}}
\begin{document}

\title{Use of Uncertain Additional Information \\ in Newsvendor Models}

\author{\IEEEauthorblockN{1\textsuperscript{st} Sergey Tarima}
\IEEEauthorblockA{\textit{Institute for Health and Equity} \\
\textit{Medical College of Wisconsin}\\
Wauwatosa, U.S.A. \\
starima@mcw.edu}
\and
\IEEEauthorblockN{2\textsuperscript{nd} Zhanna Zenkova}
\IEEEauthorblockA{\textit{Institute of Applied Mathematics and Computer Science} \\
\textit{Tomsk State University}\\
Tomsk, Russia \\
zhanna.zenkova@mail.tsu.ru}
}

\maketitle

\begin{abstract}
The newsvendor problem is a popular inventory management problem in supply chain management and logistics. Solutions to the newsvendor problem determine optimal inventory levels. This model is typically fully determined by a purchase and sale prices and a distribution of random market demand. From a statistical point of view, this problem is often considered as a quantile estimation of a critical fractile which maximizes anticipated profit. The distribution of demand is a random variable and is often estimated on historic data. In an ideal situation, when the probability distribution of the demand is known, one can determine the quantile of a critical fractile minimizing a particular loss function. Since maximum likelihood estimation is asymptotically efficient, under certain regularity assumptions, the maximum likelihood estimators are used for the quantile estimation problem. Then, the Cramer-Rao lower bound determines the lowest possible asymptotic variance. Can one find a quantile estimate with a smaller variance then the Cramer-Rao lower bound? If a relevant additional information is available then the answer is yes. Additional information may be available in different forms. This manuscript considers minimum variance and minimum mean squared error estimation for incorporating additional information for estimating optimal inventory levels. By a more precise assessment of optimal inventory levels, we maximize expected profit.
\end{abstract}


\section{Introduction}

Newsvendor model is a popular inventory management model. This model depends on two simple quantities, the purchase price of a single unit of a product $c$ and the its price when it is sold, $p$. The overall profit is fully defined by the critical fractile $(=(p-c)/p)$. If $F$ is the distribution of the random demand, $D$, then the optimal amount of product to order is $Q = F^{-1}\left((p-c)/p\right)$, where $F^{-1}$ is an inverse of $F$. This solution secures the highest expected profit.

The distribution of the random variable $D$ is often estimated using historic data. Hayes \cite{Hayes1969} considered exponential and Gaussian models to minimize Expected Total Operating Cost in the newsvendor problem. A Bayesian approach was used to improve the estimation. Similarly, Bayesian methodology was used to solve inventory problems in \cite{Scarf1959} and \cite{Karlin1960}. In more recent literature Bayesian frameworks is also very popular.

Liyanagea and Shanthikumar \cite{Liyanagea2005} suggested to use direct profit maximization which simultaneously incorporates both parameter estimation and expected profit maximization. 

Quantile estimation methods range from maximum likelihood estimators to more robust methods minimizing specific risk functions: Koenker's quantile regression \cite{Koenker2005} minimizes the sum of absolute deviations of residuals; sum of signs of residuals minimization is suggested in \cite{Tarassenko2014}. All of these methods directly applicable to solving the newsvendor problem as well. 

In a situation, when a family of probability distributions of the demand is known, a maximum likelihood estimation can be applied to quantile estimation as maximum likelihood estimators (MLEs) are asymptotically efficient under certain regularity assumptions. Then, asymptotically, MLEs reach the Cramer-Rao lower bound for variance, and no other estimator has asymptotically smaller variance. Can a quantile estimator with a smaller variance than the Cramer-Rao lower bound be found? If a relevant additional information is available then the answer is yes.

Many of the above mentioned statistical methods to solving the newsvender problem lead to estimators regular enough to have two finite moments. 
If the two moments exist, then additional information can be combined together with external information (for example, an averaged sales from another store with similar characteristics) known with a degree of uncertainty (for example, a standard error of this average is known as well) \cite{Tarima2006}. This approach assumed that the additional information is unbiased, meaning that averaged sales for both stores are about the same. A similar assumption was made in \cite{Tarima2007}. It is possible that additional information can be biased, then minimum mean squared error (MSE) can be considered instead \cite{Tarima2009, Tarima2013}. The use of additional information known up to a few distinct values is considered in \cite{Dmitriev2014,Dmitriev2015, Dmitriev2017, Dmitriev2017_2}; the minimum MSE criterion was also used in these papers. Zenkova and Krainova \cite{Zenkova2017} considered the use of a known quantile for estimating expectations. The net premium using a known quantile for voluntary health insurance was used as an illustrative application.

This manuscript considers minimum variance and minimum MSE estimation for incorporating additional information. Section \ref{Methodology} presents methodology for combining empirical data (historical sales data directly available for data analysis) and external information available in form of means and standard errors. Sections \ref{Illustrative_Example1} and \ref{Illustrative_Example2} illustrate the use of this new statistical approach to the newsvendor problem for quantile estimation.

\section{Illustrative Example}\label{Illustrative_Example1}
Table \ref{tab:1} reports an artificial dataset with $36$ weeks sales data. Product A is sold at $860$ dollars per unit, Product  B  is  sold  at  $490$ dollars  per  unit.  The  retailer  pays $660$ dollars/unit for product A and $370$ dollars/unit for product B.

\begin{table}[ht]
    \centering
    \begin{tabular}{c|c|cccccc}
        Product A &  &  &  &  &  & \\
       \hline
         6576 & 4263 & 5340 & 3697 & 3535 & 2651\\
         2541 & 2351 & 3611 & 3867 & 4257 & 6204\\
         6666 & 4364 & 5441 & 3727 & 3495 & 2755\\
         2399 & 2452 & 3621 & 3961 & 4291 & 6264\\
         6600 & 4333 & 5391 & 3732 & 3662 & 2498\\
         2576 & 2402 & 3588 & 3900 & 4220 & 6214\\
       \hline
        Product B &  &  &  &  &  & \\
       \hline
        215 & 142 & 155 & 97 & 101 & 83\\
        104 & 96 & 102 & 101 & 130 & 215\\
        223 & 134 & 157 & 99 & 99 & 87\\
        100 & 97 & 98 & 104 & 131 & 202\\
        211 & 139 & 150 & 100 & 105 & 82\\
        103 & 98 & 100 & 102 & 127 & 219
    \end{tabular}
    \caption{thirty six week sales history (in numbers of units sold) for Products A and B}
    \label{tab:1}
\end{table}
The retailer is mainly interested in Product A as it is associated with high sales and is highly  important for the retailer's success. Then, the critical fractile ratio for Product A is $23.26\%(=(860-660)/860)$. The solution to the newsvendor problem is $Q=F^{-1}(0.2326)$. Relying on previous experience the retailer is confident that market demand for the two products, A and B, can be described by normal distributions, and the demands are likely to be correlated. Seasonal variation is so small that historic weekly data can be assumed to be independent. The only complication is that there exists a difficult to predict clustering, see Figure \ref{fig1}, but this problme is not in the focus of this manuscript. Normal distributions depend on two unknown parameters: $\mu$ and $\sigma^2$. Further, we will use subscripts $A$ and $B$ to differentiate between $\mu$ and $\sigma^2$ of Products A and B when needed.

Using Table \ref{tab:1}, the MLEs of the unknown parameters of the normal model are $\hat\mu_A = 4095.694$ (sample mean) and $\hat\sigma^2_A = 1791703$ (sample variance). Note, the sample variance is slightly different from the MLE of $\sigma_A^2$, but the sample variance is an unbiased estimator of $\sigma_A^2$ and will be used instead.

Then, using the $0.2326$-level normal quantile, $Q = 3118.14$ (in Product A units). This is approximately the MLE of optimal inventory levels for Product A. It is possible that another estimating procedure can be chosen to evaluate $Q$. For example, a direct quantile estimation without any assumptions on the underlying parametric family will lead to another estimate = $2859.34$. For illustrative purposes, we will focus on the approximate MLE and assume that the estimate is approximately unbiased. Without loss of generality the approach on the use of additional information, considered in this paper, applies to all unbiased or approximately unbiased estimators with two finite moments.

Given the importance of Product A, it is very difficult to obtain historic sales data from similar retailers. At the same time, additional data on Product B is much simpler to get from other retailers. Since the information on Product B is viewed as less important by other retailers, they freely share their sales data over a cup of coffee. 

Consider the following additional information. An owner of a similar retailer store bragged that his store sold $30$ thousand units of Product B in past five years (additional information $1$), whereas an owner of another similar store said that his sales of Product B are higher than $100$ units every other week within the same five year period (additional information $2$). Can we use these two pieces of seemingly irrelevant information to improve estimation accuracy of the optimal inventory levels for Product A? The answer is yes, and we will return to this illustrative example in Section \ref{Illustrative_Example2}.

\section{Methodology}\label{Methodology}
Let $\theta$ be the parameter of interest, $\theta = F^{-1}((p-c)/p)$ for the newsvendor problem. The estimator of $\theta$ based on historical data, $\hat\theta$, is assumed to be an unbiased estimator of $\theta$, so that $E(\hat\theta) = \theta$. The $\hat\theta$ is a normal quantile estimated on historical sales data in Section \ref{Illustrative_Example1} ($\hat\theta = 3118.14$). In addition to $\hat\theta$, another estimator $\tilde\eta$ is available as additional information. This quantity estimates $\eta$ not $\theta$, and $\eta$ is a different and possibly multi-dimensional parameter. Specifically, in Section \ref{Illustrative_Example1}, $\eta=\left(\eta_{1},\eta_{2}\right)$, where $\eta_1$ is the mean weakly sales of Product B and $\eta_2$ is the median weakly sales of Product B. The additional information described in Section \ref{Illustrative_Example1} can be converted into a two-dimensional estimate $\left(\tilde\eta = (115.3846,100)\right)$. The number $115.3846$ is obtained as a ratio $30,000/260$, because there are $260$ weeks within a five year period. The first additional information sets the mean weakly sales of Product B at $115.3846$ units, and the second additional information sets the median sales at $100$ units/month. 

Further, we use ``hat'' to denote estimators based on empirical (historical) data and ``tilde'' for the quantities determined by additional information.
Using the data in Table \ref{tab:1}, the mean weakly sales = $128$ ($\hat\eta_1 = 128$) and weakly median sales = $103.2$ ($\hat\eta_2$).

To combine additional information with empirical data, we consider a class of linear combinations 
\begin{equation}
\theta^{\Lambda} = \hat\theta + \Lambda\left(\hat\eta - \tilde\eta\right)
\end{equation} 
In the above equation, $\hat\eta$ refers to an estimate of $\eta$ based on empirical data. It is clear that $E(\hat\eta) = \eta$ by a property of the sample mean and median of normal data, but $E(\tilde\eta) = \eta + \delta$, where $\delta$ is a possible bias (a vector-column of biases) associated with additional information. In Section \ref{Illustrative_Example1}, the bias has two components and
$$\hat\delta = \hat\eta - \tilde\eta = (12.6154, 3.5).$$

Following \cite{Tarima2009}, the smallest MSE in the class $\theta^{\Lambda}$ is secured with
\begin{equation}
\label{optest_MMSE_mult} 
\theta^{0}(\delta) = \hat\theta - cov\left(\hat\theta, \hat\delta\right)E^{-1}\left(\hat\delta\hat\delta^T\right) \hat\delta^T
\end{equation}
and
\begin{eqnarray}
\label{MSEoptest_MMSE_mult} 
MSE\left(\theta^{0}\right) = cov\left(\hat\theta\right) - 
cov\left(\hat\theta, \hat\delta\right)E^{-1}\left(\hat\delta\hat\delta^T\right)cov\left(\hat\delta, \hat\theta\right),
\notag
\end{eqnarray}
where $E\left(\hat\delta\hat\delta^T\right) = cov\left(\hat\eta\right) + cov\left(\tilde\eta\right) + \delta\delta^T$. 

The special case of $\delta = 0$ makes $\theta^{\Lambda}$ unbiased for all choices of $\Lambda$. Then, 
\begin{equation}
\label{optest_MVAR_mult} 
\theta^{0}(0) = \hat\theta - cov\left(\hat\theta, \hat\delta\right)cov^{-1}\left(\hat\delta\right) \hat\delta^T
\end{equation}
has the smallest variance among all $\theta^{\Lambda}$, see \cite{Tarima2006}, and 
\begin{eqnarray}
\label{MSEoptest_MVAR_mult} 
cov\left(\theta^{0}(0)\right) = cov\left(\hat\theta\right) - 
cov\left(\hat\theta, \hat\delta\right)cov^{-1}\left(\hat\delta\right)cov\left(\hat\delta, \hat\theta\right).
\notag \end{eqnarray}
For one-dimensional $\theta$, the quadratic form on the right hand side of Equation \ref{MSEoptest_MMSE_mult} 
$$M = cov\left(\hat\theta, \hat\delta\right)cov^{-1}\left(\hat\delta\right)cov\left(\hat\delta, \hat\theta\right)\ge 0.$$
Two extreme scenarios, associated with $M$, describe how relevant additional information is for estimating $\theta$:
\begin{itemize}
\item If $\hat\theta$ and $\hat\delta$ are uncorrelated, then $M=0$ and $\theta^{0}(\delta) = \hat\theta$ for all $\tilde\eta$.     
\item If $cov\left(\hat\theta, \hat\delta\right)=cov\left(\hat\theta\right)$ (the case of exact knowledge $\tilde\eta = \theta$) then 
$M=cov\left(\hat\theta\right)$, and $\theta^{0}(0) = \theta$ and $cov(\theta^{0}(0))=0$.   
\end{itemize}

The $\theta^{0}(\delta)$ is not directly applicable in practice as covariances are unknown. In addition, the unknown $\delta$ is also present in its structure. Dmitriev and his colleagues \cite{Dmitriev2014} explored the same class of estimators. In contrast to our settings, they hypothesized that  $\tilde\eta=\eta+\delta$ is known to belong to a distinct set of pre-determined values.

We estimate unknown covariances on empirical data to obtain an approximation to the optimal $\theta^0(\delta)$:
\begin{equation}
\label{hat_MMSE_mult2} 
\hat\theta^{0}(\delta) = \hat\theta - \widehat{cov}\left(\hat\theta, \hat\delta\right)
\left(\widehat{cov}\left(\hat\eta\right) + \widetilde{cov}\left(\tilde\eta\right) + \delta\delta^T\right)^{-1}
\hat\delta^T
\end{equation}
The useful property of $\hat\theta^{0}(\delta)$ is that it is easy to show 
that under some regularity conditions 
\begin{equation}
\sqrt{n}\left(\hat\theta^{0}(\delta)-\theta^{0}(\delta)\right) = o_p(1).\label{convergence0}
\end{equation}
Another interesting asymptotic result is that $\forall\delta \ne 0$, 
\begin{equation}\label{conv1}\sqrt{n}\left(\theta^{0}(\delta) - \theta\right) = o_p(1)\end{equation} and 
\begin{equation}\label{conv2}\sqrt{n}\left(\hat\theta^{0}(\delta) - \theta\right) = o_p(1).\end{equation}
From (\ref{conv1}) and (\ref{conv2})
\begin{equation}\label{conv3}\sqrt{n}\left(\hat\theta^{0}(\delta) - \theta^{0}(\delta)\right) = o_p(1).\end{equation}

Estimator $\hat\theta^{0}(\delta)$, however, still includes an unknown $\delta$. If we plug in $\hat\delta$
instead, we get another approximation:
\begin{equation}
\label{hat_MMSE_mult3} 
\hat\theta^{0}(\hat\delta) = \hat\theta - \widehat{cov}\left(\hat\theta, \hat\delta\right)
\left(\widehat{cov}\left(\hat\eta\right) + \widetilde{cov}\left(\tilde\eta\right) + \hat\delta\hat\delta^T\right)^{-1}
\hat\delta^T.
\end{equation}

The use of $\hat\delta$ in (\ref{hat_MMSE_mult2}) creates certain difficulties for (\ref{convergence0}) to hold. 
Specifically, if $\delta=0$ then $\sqrt{n}\left(\hat\theta^{0}(\hat\delta) - \theta^{0}(0)\right) = O_p(1)$, meaning that $\sqrt{n}\left(\hat\theta^{0}(\hat\delta) - \theta^{0}(0)\right)$ does not converge to zero, in probability, as $n\to\infty$; even asymptotically it continues to be a non-degenerate random variable. 

Overall, if $\delta=0$ can be surely assumed, minimum variance estimator $\hat\theta^{0}(0)$ is to be used, and if some protection against possible bias (disinformation) is needed minimum MSE estimation with $\hat\theta^{0}(\hat\delta)$ is a better choice with the understanding that $\hat\theta^{0}(\hat\delta)$ is inferior to $\hat\theta^{0}(0)$ under $\delta=0$. The estimator $\hat\theta^{0}(\delta)$ can be used to evaluate the impact of bias on the estimating procedure. 

\section{Illustrative Example (continuation)} 
\label{Illustrative_Example2}
As it was shown in Section \ref{Methodology} a minimum variance estimator $\hat\theta^{0}(0)$ and a minimum MSE estimator $\hat\theta^{0}(\hat\delta)$  are the estimators to use in practice. This section shows how both estimators can be calculated using R package ``AddInf'' available at {\begin{verbatim}https://github.com/starima74/AddInf\end{verbatim}}  

\begin{figure}
    \centering
    \includegraphics[scale=0.44]{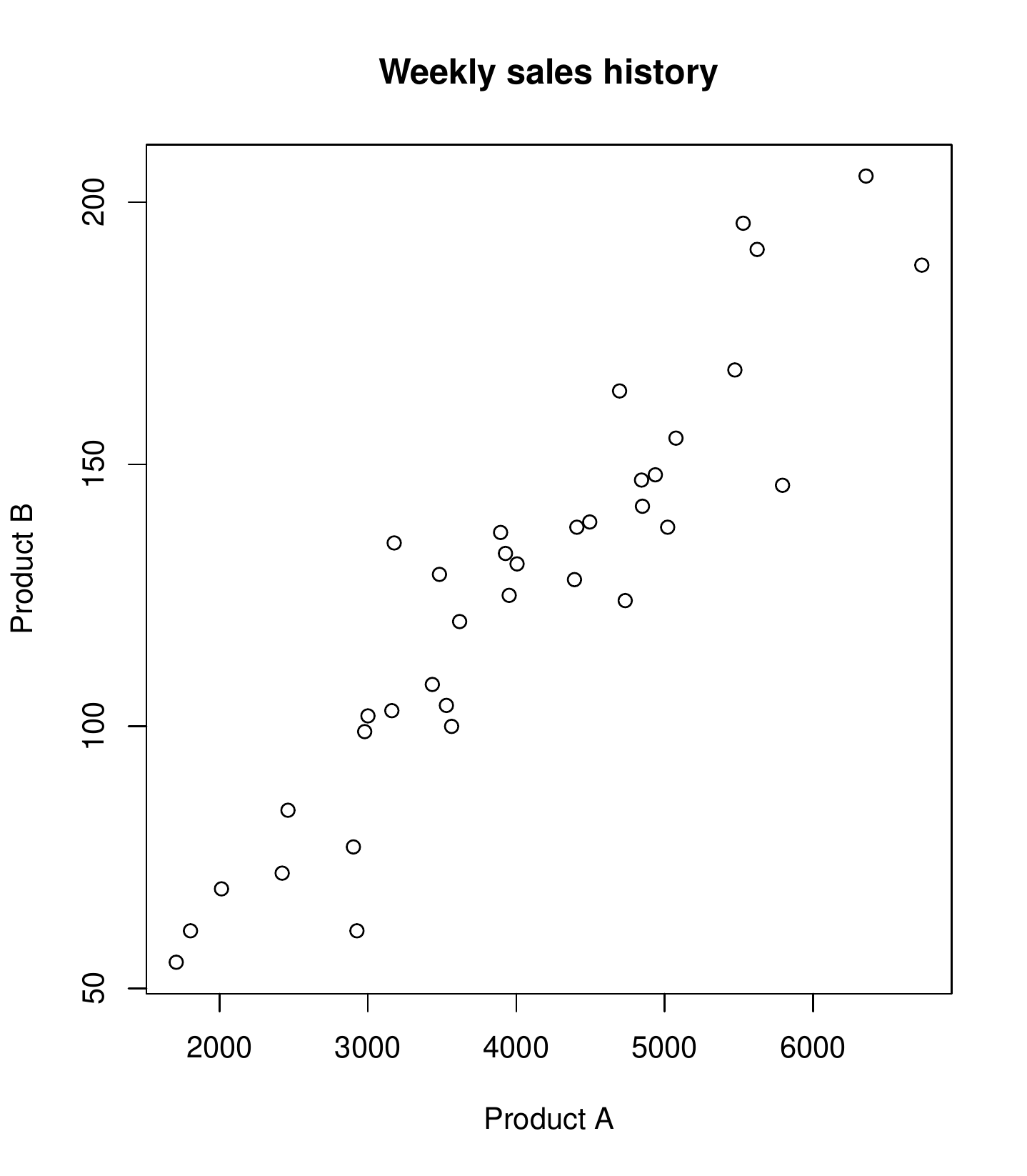}
    \caption{Scatter plot of sales of Products A and B; Pearson correlation is 93.6\%.}
    \label{fig1}
\end{figure}
To use ``AddInf'' R package one need to install ``devtools'' library use its ``install\_github'' command to install ``AddInf'' package
\begin{verbatim}
install.packages("devtools")
library(devtools)
install_github("starima74/AddInf", 
        force=TRUE)
library(AddInf)    
\end{verbatim}

The following part of R code creates weekly sales data for products A and B. 
\begin{verbatim}
A <- c(6576,	4263,	5340,	3697,	3535,	2651,
       2541,	2351,	3611,	3867,	4257,	6204,
       6666,	4364,	5441,	3727,	3495,	2755,
       2399,	2452,	3621,	3961,	4291,	6264,
       6600,	4333,	5391,	3732,	3662,	2498,
       2576,	2402,	3588,	3900,	4220,	6214)

B <- c(215, 142, 155,	97,  101, 83,	
       104, 96,  102,	101, 130, 215,
       223, 134, 157,	99,  99, 87,	
       100, 97,  98,	104, 131, 202,
       211, 139, 150,	100,  105, 82,	
       103, 98,  100,	102, 127, 219)
dd <- data.frame(A = A, B = B)

plot(dd, ylab="Product B",
     xlab="Product A", 
     main="Weekly sales history")

cor(dd)
## corr = 0.936
\end{verbatim}
First of all, distributions of sales of Products A and B need to be explored, see Figure \ref{fig1}. One can easily conclude that the association between the sales of Products A and B is linear and strong, which is supported by the Pearson correlation of $93.6\% (p<0.0001)$.

The R function below estimates optimal inventory levels at the critical fractile equal to 0.2326. The data in the argument is assumed to follow a normal distribution.
\begin{verbatim}
theta.f <- function(d) {
   qnorm(0.2326, mean = mean(d$A), 
   sd = sd(d$A))
}
\end{verbatim}

\subsection{Minimum Variance Estimation}
Additional information is aggregated into a data frame in form of means ($\tilde\eta$) and covariances ($cov(\tilde\eta)$). It is important to be able to estimate $\eta$ on the empirical data. This is why the function implementing estimation of $\eta$ needs to be defined. Here, functions are statistical procedures to calculate $\hat\eta$ using empirical data. We will discuss how to incorporate possible biases in Section \ref{Biases}
\begin{verbatim}
### (Empty Lists) to save 
### additional Information
Add.Info.Means <- list()
Add.Info.Vars <- list()
Add.Info.Functions <- list()
Add.Info.Biases <- list()
\end{verbatim}

The first additional data source declares weekly sales of Product B at $115.3846$ units. Since $var(\tilde\eta_1)$ is not available as additional information, but it is stated that additional information comes from a similar store. We can realistically assume that variance is similar as well, $var(\tilde\eta_1)=var(\hat\eta_1)$. Sample variance based on $36$ observations for Product B is $1912.8$. Then, $var(\tilde\eta_1)$ is approximated by $1912.8/260$. Since additional information is given by an averaged value, the function is simply the sample average.

\begin{verbatim}
Add.Info.Means[[1]] <- 115.3846
Add.Info.Vars[[1]] <- 1912.8/260
Add.Info.Functions[[1]] <- 
  function(d) mean(d$B,na.rm = TRUE)
\end{verbatim}

Information from the second data source is summarized into lists in a similar manner. The second source reported median = $100$. To estimate its variance,  the sample on Product B is bootstrapped as follows.
\begin{verbatim}
### variance of MEDIAN 
### (bootstrapping)
set.seed(123)
res <- 1:10000 
for(i in 1:10000) 
  res[i] <- 
  median(B[sample(1:36,replace=TRUE)])
var(res)*36
### 3227.319
\end{verbatim}
Then, $3227.319$ is used to estimate variance of $\tilde\eta_2$ with $3227.319/260$. Further,
\begin{verbatim}
Add.Info.Means[[2]] <- 100
Add.Info.Vars[[2]] <- 3227.319/260
Add.Info.Functions[[2]] <- 
  function(d) median(d$B, na.rm = TRUE)
\end{verbatim}
Note, that in the above R code $\eta_2$ is defined as the median. The lists with additional information (means, variances and functions) are aggregated into a single data frame
\begin{verbatim}
Add.Info <- data.frame(
     Means = rep(NA,2), 
     Vars = rep(NA,2), 
     Functions = rep(NA,2),
     Biases = rep(NA,2))
Add.Info$Means = Add.Info.Means
Add.Info$Vars = Add.Info.Vars
Add.Info$Functions = Add.Info.Functions
\end{verbatim}

Finally, we run ``MVAR'' function where minimum variance with additional information is implemented. This function internally uses non-parametric bootstrap to estimate unknown covariances needed for $\hat\theta^0(0)$. In addition to the first three arguments described above, the function also  uses number of bootstrap resamples (nboot) and a cutoff on the proportion of eigenvalues, which is a convenient way to deal with weakly definite covariance matrices. 
\begin{verbatim}
res <- MVAR(dd, theta.f, Add.Info, 
   nboots = 5000, eig.cutoff = 1)
\end{verbatim}
The result consists of $\hat\theta^0(0)$ and its variance, and $\hat\theta$ and its variance. 
\begin{verbatim}
res
### $`Theta.Est`
###         [,1]
### [1,] 3072.728
###
### $Theta.Est.Var
###         [,1]
### [1,] 904.6197
###
### $Theta.Hat
### [1] 3118.14
###
### $Theta.Hat.Var
### [1] 1060.981
\end{verbatim}
From these results we can estimate the standard deviation of $\hat\theta^0(0)$ which is approximately equal to $30.0769 (=\sqrt{904.6197})$, and the standard deviation of $\hat\theta$, which is $32.5727 (=\sqrt{1060.981})$. Thus, the asymptotic confidence interval becomes 8\% shorter with the use of additional information. Taking into account that the covariances are estimated with bootstrap resampling, the results may slightly differ from one run to another. To avoid this randomness, the ``set.seed'' function can be used:
\begin{verbatim}
set.seed(123)
\end{verbatim}
The above improvement in standard deviations may seem marginal, but for other choices of additional information the changes can be more visual. For example, what if we assume that the variance of the first additional information is much smaller (for example, the 5 year sales of $30,000$ units of Product B is an averaged value across 10 different stores, which corresponds to $2600$ weeks of follow-up):
\begin{verbatim}
Add.Info.Vars[[1]] <- 1912.8/2600
Add.Info$Vars = Add.Info.Vars
\end{verbatim}
After running the MVAR function
\begin{verbatim}
res <- MVAR(dd, theta.f1, Add.Info, 
   nboots = 5000, eig.cutoff = 1)
\end{verbatim}
the result is
\begin{verbatim}
res
### $`Theta.Est`
###          [,1]
### [1,] 2962.054
###
### $Theta.Est.Var
###          [,1]
### [1,] 629.5974
###
### $Theta.Hat
### [1] 3118.14
### 
### $Theta.Hat.Var
### [1] 998.8053
\end{verbatim}
From the above, the standard deviation of $\hat\theta^0(0)$
is $25.09178$ and the SD of $\hat\theta$ is $31.60388$ leading to a $21\%$ reduction in the width of the asymptotic confidence interval.

\subsection{Minimum Mean Square Estimation}\label{Biases}
It is not impossible that the additional information came from a biased source. For example, what if the additional information said that 5 year sales were equal to $100,000$ units of Product B, not $30,000$? This leads to a very different estimated weekly sales of $384.6154$ units of Product B ($=100,000/260$). 
\begin{verbatim}
Add.Info.Means[[1]] <- 100000/260
Add.Info$Means = Add.Info.Means
\end{verbatim}
If this information is incorrect, it will lead to a very different and misleading optimal inventory levels for product A:
\begin{verbatim}
### $`Theta.Est`
###        [,1]
### [1,] 6238.16
###
### $Theta.Est.Var
###         [,1]
### [1,] 635.5988
###
### $Theta.Hat
### [1] 3118.14
###
### $Theta.Hat.Var
### [1] 1002.272
\end{verbatim}
This, incorrect or deliberately altered additional information may lead to serious biases and, thus, have to be approaches with extreme caution. On the other hand, if such additional information is actually correct, it can fix optimal inventory assessment based on low quality empirical data.

If, however, the empirical data is of good quality, the minimum MSE approach can be used, which provides robustness to undue influence of additional information. In this case, an indicator of possible bias needs to be added to the ``Add.Info'' data structure:
\begin{verbatim}
Add.Info.Biases[[1]] <- 1
Add.Info.Biases[[2]] <- 0
Add.Info$Biases = Add.Info.Biases
\end{verbatim}
The above R code defines that the first source of additional information may be unreliable, whereas the second source is reporting unbiased additional information. In this situation,  $\hat\theta^0(\hat\delta)$ should be applied instead
\begin{verbatim}
res <- MMSE(dd, theta.f, Add.Info, 
   nboots = 5000, eig.cutoff = 1)
res
###  $`Theta.Est`
###          [,1]
### [1,] 3110.567
###
### $Theta.Est.Var
###         [,1]
### [1,] 976.0103
###
### $Theta.Hat
### [1] 3118.14
###
### $Theta.Hat.Var
### [1] 1045.683
\end{verbatim}
Now, the $\hat\theta^0(\hat\delta)=3110.567$ and is very close to 
$\hat\theta=3118.14$. The unreliable additional information is automatically suppressed, as its serious bias is easily detected, whereas the second source still contributed to a smaller asymptotic variance $976.0103$ of $\hat\theta^0(\hat\delta)$ versus $1045.683$ of $\hat\theta$.

If the bias is small, for example ($\tilde\eta_1 = 130$ and $\hat\eta_1 = 128$)
\begin{verbatim}
Add.Info.Means[[1]] <- 130
Add.Info$Means = Add.Info.Means
\end{verbatim}
then 
\begin{verbatim}
res <- MMSE(dd, theta.f, Add.Info, 
    nboots = 5000, eig.cutoff = 1)
res
### $`Theta.Est`
###         [,1]
### [1,] 3114.023
###
### $Theta.Est.Var
###         [,1]
### [1,] 647.4465
###
### $Theta.Hat
### [1] 3118.14
###
### $Theta.Hat.Var
### [1] 1002.441
\end{verbatim}
Thus, when additional information is consistent with empirical data, MMSE and minimum variance approaches show similar improvement of asymptotic variance.

\section{Summary}\label{Summary}

Additional information in form of known statistical quantities and their standard errors can be helpful in estimating many statistical quantities including optimal inventory levels in the newsvendor models. This manuscript shows how to incorporate such additional information into statistical estimation. An illustrative example on estimating optimal inventory levels with additional information is analyzed with the R package ``AddInf''. 

The illustrative example shows how information from multiple additional sources can be used. If the additional information is not correct or deliberately altered (disinformation) the minimum variance estimation may be inappropriate. At the same time minimum mean squared estimation detects that such additional information is inconsistent with the empirical data and its impact is automatically suppressed. On the other hand if the additional information is consistent with the empirical data, the benefits of using minimum variance and minimum MSE approach are comparable. 

Thus, with the use of additional information, more accurate assessment of optimal inventory  levels is obtained, which maximizes expected profit.

\bibliographystyle{IEEEtran}
\bibliography{bib}

\end{document}